\documentclass{article}[12pt]
\usepackage{everypage}
\usepackage{authblk}
\usepackage{graphicx}
\usepackage{etex}
\usepackage{float,verbatim}
\usepackage{color,ulem}
\usepackage{amssymb,amsmath}
\RequirePackage{xspace}
\def\be{\begin{equation}}
\def\ee{\end{equation}}
\def\bee{\begin{eqnarray}}
\def\eee{\end{eqnarray}}
\def\cm {\ensuremath{{\rm \,cm}}\xspace}

\begin{document}
\title{ 
Measuring Propagation Speed  of  Coulomb  Fields}
{\small
\author{ R. de Sangro, G. Finocchiaro, P.Patteri, M. Piccolo, G. Pizzella}
\affil{\small Istituto Nazionale di Fisica Nucleare,Laboratori Nazionali 
di  Frascati}
\maketitle
{\small\date}

\abstract {The problem of gravity propagation has been subject of discussion 
for quite a long time: Newton, Laplace and, in
relatively more modern times, Eddington pointed out that, if gravity 
propagated with finite velocity, planet motion around the sun
would become unstable due to a torque originating from time lag of the 
gravitational interactions.

Such an odd behavior can be found  also in electromagnetism, when one computes
the propagation of the electric fields 
generated by a set  of uniformly moving charges.
As a matter of fact the Li\'enard-Weichert retarded potential leads to
the same formula as the one obtained 
assuming that 
the electric field propagates with infinite velocity. 
The Feynman explanation for this apparent paradox was based on the fact that 
uniform motions last indefinitely.

To verify such an explanation, we performed an experiment to measure 
the time/space evolution of the electric field
generated by an uniformly moving electron beam. The results we obtain, on 
a finite lifetime kinematical state,
are compatible with an electric field rigidly carried by the beam itself.} 

\section{Introduction}

 In  \textit{Space, Time and Gravitation} Eddington
discusses\,\cite{eddi} the problem of  gravity propagation. He remarks
that if gravity propagated with finite velocity the motion of the planets around 
the Sun would become unstable, due to a torque acting on the
planets. The problem was already  known to Newton and was examined by
Laplace\,\cite{laplace}, who calculated a lower limit for the  gravity
propagation velocity, finding a value much larger than the speed of light.

However, at the time of Eddington's writing, General
Relativity had been just formulated, with gravitational waves
traveling with the speed of light as a possible solution.
Eddington noted that a similar problem existed in
electromagnetism, and since electromagnetic waves in
vacuum do travel with the speed of light, he concluded that in General
Relativity gravity also propagates with the speed of light.

We remark that an intriguing behavior of electromagnetism occurs when
the field of an electric charge moving with constant  velocity is computed.
One finds that in such a case the electric field at a given point
$\pmb{P(x,y,z,t)}$ evaluated with the Li\'enard-Wiechert (L.W.) potentials
is identical to that calculated by assuming that the Coulomb
field travels with infinite velocity.

This is a direct consequence of the velocity field, the part of
the L.W. potentials independent of the charge acceleration, being a 
static field. This feature has been strassed by several authors 
{\it e.g.} \cite{becker} \cite{jackson}.
The Feynman \cite{feyn} interpretation is based instead on the 
assumption that the uniform motion lasts indefinitely and that 
an observer would see an angular acceleration of the approaching
charge.
 
The only way to shed some  light on this
 problem, either the Feynman interpretation or  the
static Coulomb field carried rigidly by the charge is by means of 
an experiment.
 
To verify if the Feynman interpretation of the L.W. potentials holds in  
case of a charge moving with constant velocity for a finite time,
we have  performed an experiment to measure  the time evolution
of the  electric field  produced by an electron beam in our
laboratory; such  kinematic state has obviously a finite lifetime.

It is well known that a sizable number of  instrumentation devices
(e.g. beam position monitors) 
are based on effects produced by electric fields carried by particle beams. 
The effects and the propagation of such fields, however, have never been
studied in details: the main point exploited by these devices is  that the 
field effects are  contemporary to the particles passage and that the signal
size obtained, for instance, on a pair of strip-lines inside a vacuum pipe
yields a measurement of the transverse position of the beam itself.
The experimental situation for those devices is  quite complicated 
to understand, as all the
fields are  inside a conductor and the transverse distances exploited are
always small.
We, on the contrary, tried to carry out our experiment in a clean environment:
the electron beams used were propagating in a {\it vacuum like} environment.
We covered  a wide range of transverse distances  w.r.t. the
beam line (up to 55 cm). Such range leads to explore time and space domains
for the {\it emission} of the detected field far outside the physical region.  
 
The data we collected, as shown in the following, are compatible  with the
hypothesis of a Coulonb filed carried rigidly by the moving charge

\section{Theoretical Considerations}

The electric field at ${\pmb r}(x,y,z)$ from a charge
$e$ traveling with constant velocity $\pmb v$, at a time $t$   
can be written,  using the Li\'enard-Wiechert retarded
potentials as\,\cite{landau,becker,jackson}:
\bee
{\pmb E}({\pmb r},t)=\frac{e}{4\pi\epsilon_o}
\frac{1-{v^2}/{c^2}}{\left(R(t')-
\frac{{\pmb R}(t')\cdot{\pmb v}}{c}\right)^3}
\left({\pmb R}(t^\prime)-{\pmb v}\frac{R(t^\prime)}{c}\right),
\label{ee}
\eee
where
\be
 {\pmb R}(t^\prime)={\pmb r}-{\pmb v}t^\prime
\label{y}
 \ee
is the distance between the moving charge and  the space point where
one measures the field at time $t$, and
\be 
t'=t-\frac{R(t')}{c}.
\label{gg}
\ee

The field from a steadily moving charge can also be written (as easily 
deducible from Eq. \ref{ee} in case of constant velocity)
\cite{feyn,landau,becker,jackson} as
\be
{\pmb E}(t)=\frac{e}{4\pi\epsilon_o}
\frac{{\pmb R}(t)}{R(t)^3}
\frac{1-{v^2}/{c^2}}{(1-\frac{v^2}{c^2}sin^2(\theta(t))^{\frac{3}{2}}}
\label{e}
\ee
where  $\pmb R(t)$
is the vector joining the charge  position and the point at
which we evaluate the e.m. field at time $t$  (Eqs.\,38.8 and 38.9 of\,\cite{landau})\footnote{In Landau's
  words:{\it .....the distance $\pmb R(t)$ at precisely the moment of
  observation} (see pag.\,162 in \cite{landau}).} and $\theta(t)$ is the angle between 
$\pmb v$ and $\pmb R(t)$ 

A pictorial view of the above mentioned quantities can be seen in Fig.\ref{fig1}.
\begin{figure}[hbtp]
\begin{center}
\includegraphics[scale=0.35]{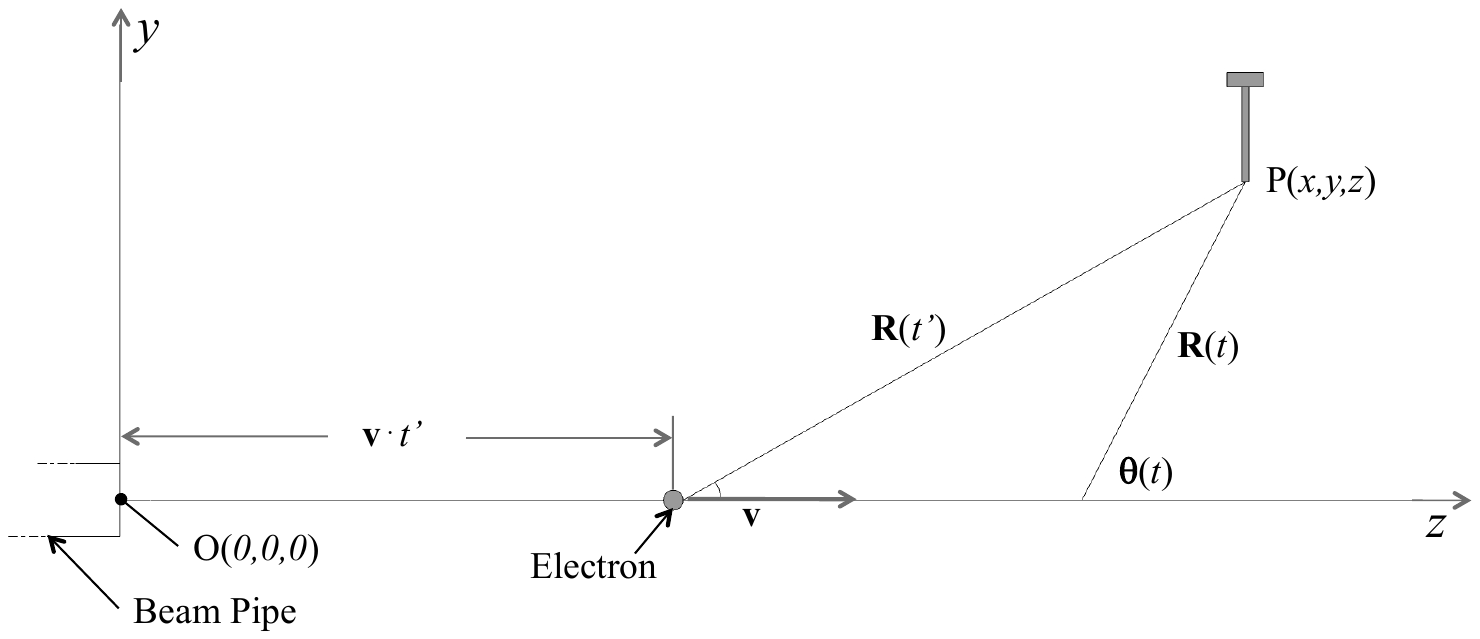}
\caption{
  A pictorial view of various quantities mentioned in eqns \ref{ee} and \ref{e}.}
\label{fig1}
\end{center}
\end{figure}
If we indicate with $y$ the generic transverse coordinate, using eqn.\ref{ee} we can compute 
the maximum transverse electric
field w.r.t. the direction of motion,
given by ($\gamma\equiv1/\sqrt{1-v^2/c^2}$):
\be
   E_{max}=\frac{e}{4\pi\epsilon_o}\frac{\gamma}{y^2},
\label{dd}
\ee
a value obtained when the charge is at a distance $\gamma y$ at a time
\be
t'=t-\frac{\gamma y}{c}
\label{tprimo}
\ee
.
\begin{figure}[hbtp]
\begin{center}
\includegraphics[scale=0.5]{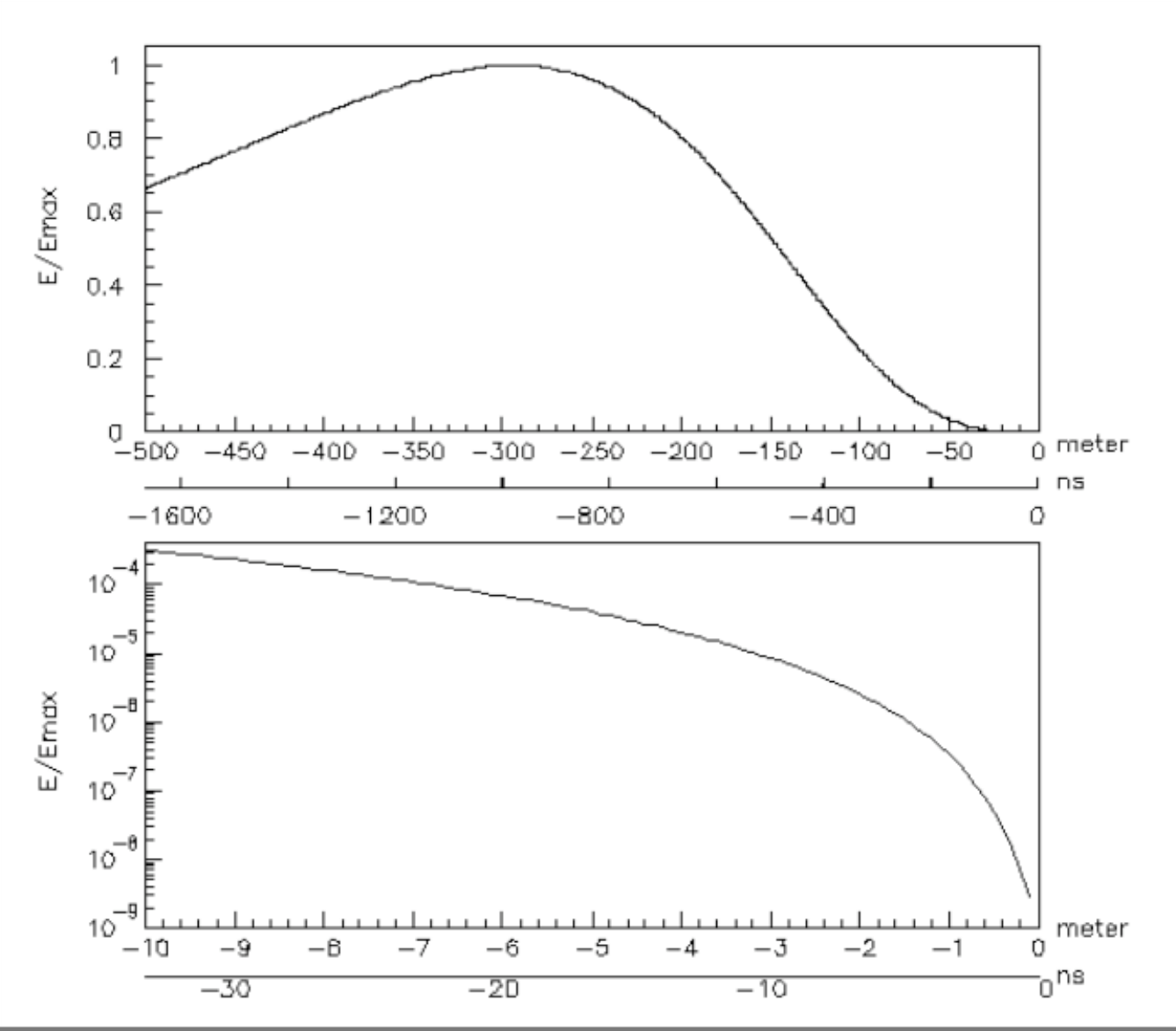}
\caption{
  The electric field from eqn.\,\ref{ee} normalized to its maximum value,
  ${E_y(R)}/{E_{max}}$, generated by 500\,MeV electrons as a function of
  $z^\prime$ (or $t^\prime$, lower abscissa scales), expected at
  $(z=0\cm, y=30\cm)$.
  $z'$ and $t'$ are defined in eqns.\,\ref{y} and \ref{gg}.  
  The horizontal scale of the upper graph (a) is such to include the
  point where ${E_y(R)}={E_{max}}$; the lower graph (b) is a close-up
  of the region $z\in [-10,0]\,$m typical of our experiment (note the
  different vertical scales).}
\label{fig2}
\end{center}
\end{figure}

 Fig.~\ref{fig2} shows the field, normalized to $E_{max}$,
generated by relativistic electrons ($E=500\,$MeV) moving along the
$z$ axis, at a transverse distance $y=30\,$cm. We
observe that the maximum value of the field appears to be generated
when the charges are in an unphysical region, namely
$z=-300\,$m. Conversely, the calculated electric field in the region of our
experiment ($|z|\le10\,$m) is many orders of magnitude smaller.

\section{The Experiment}
In our experiment we measure the electric field generated by
the electron beam produced at the DA$\Phi$NE Beam Test Facility
(BTF)~\cite{BTF}, a beam line built and operated at the Frascati National 
Laboratory to produce a well-defined number of electrons (or
positrons) with energies between 50 and 800 MeV.
At maximum intensity the facility yields, at a 50\,Hz 
repetition rate, 10 nsec long beams with a total charge  up to several
hundreds pCoulomb.
The electron beam is delivered to the 7\,m long experimental hall in a
beam pipe of about 10\,\cm diameter, closed by a $40\,\mu$m Kapton window.
Test were carried out shielding the exit window with a thin copper layer, 
but we did not observe any change in the experimental situation.
At the end of the hall a lead beam dump absorbs the beam particles.   
In our measurements we used 500\,MeV beams of $0.5\div 5.0\times 10^8$
electrons/pulse ($\gamma\simeq 10^3$).

A schematic view of the experimental setup is shown in Fig.\,\ref{fig3}.
At the beam pipe exit flange, electrons go through  a fast toroidal 
transformer  measuring total  charge and providing redundancy on 
our LINAC-RF based  trigger.

\begin{figure}
\begin{center}
\includegraphics[scale=0.35]{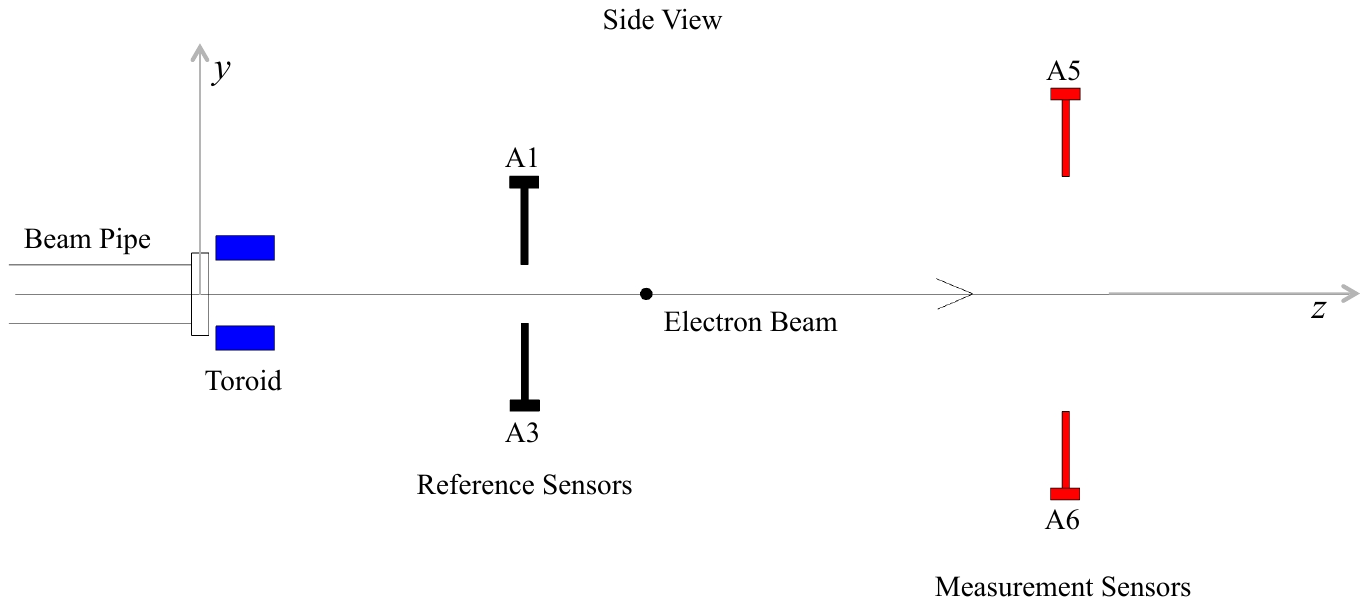}
\caption{
   A schematic side viewof our  experimental apparatus. Sensors A2 and A4, orthogonal to
the figure plane, are not shown.}
\label{fig3}
\end{center}
\end{figure}

To measure the electric field we used
as sensors 14.5\,\cm long,  0.5\,\cm diameter Copper round bars,
connected to our Data Acquisition System  by means of  fast, 
terminated coax cables.

To record the sensors waveforms we used a Switched Capacitor Array (SCA)  circuit
(CAEN mod V1472) able to sample the input signal at 5 GHz. In 
addition to the sensors output, the SCA stored also the LINAC-RF trigger 
and the toroid pulse.

The Coulomb field acts on our sensor quasi-free electrons, generating a 
current.
An example of the recorded signals is shown in Fig.\ref{fig4}.
\begin{figure}[hbtp]
\includegraphics[ scale=.7]{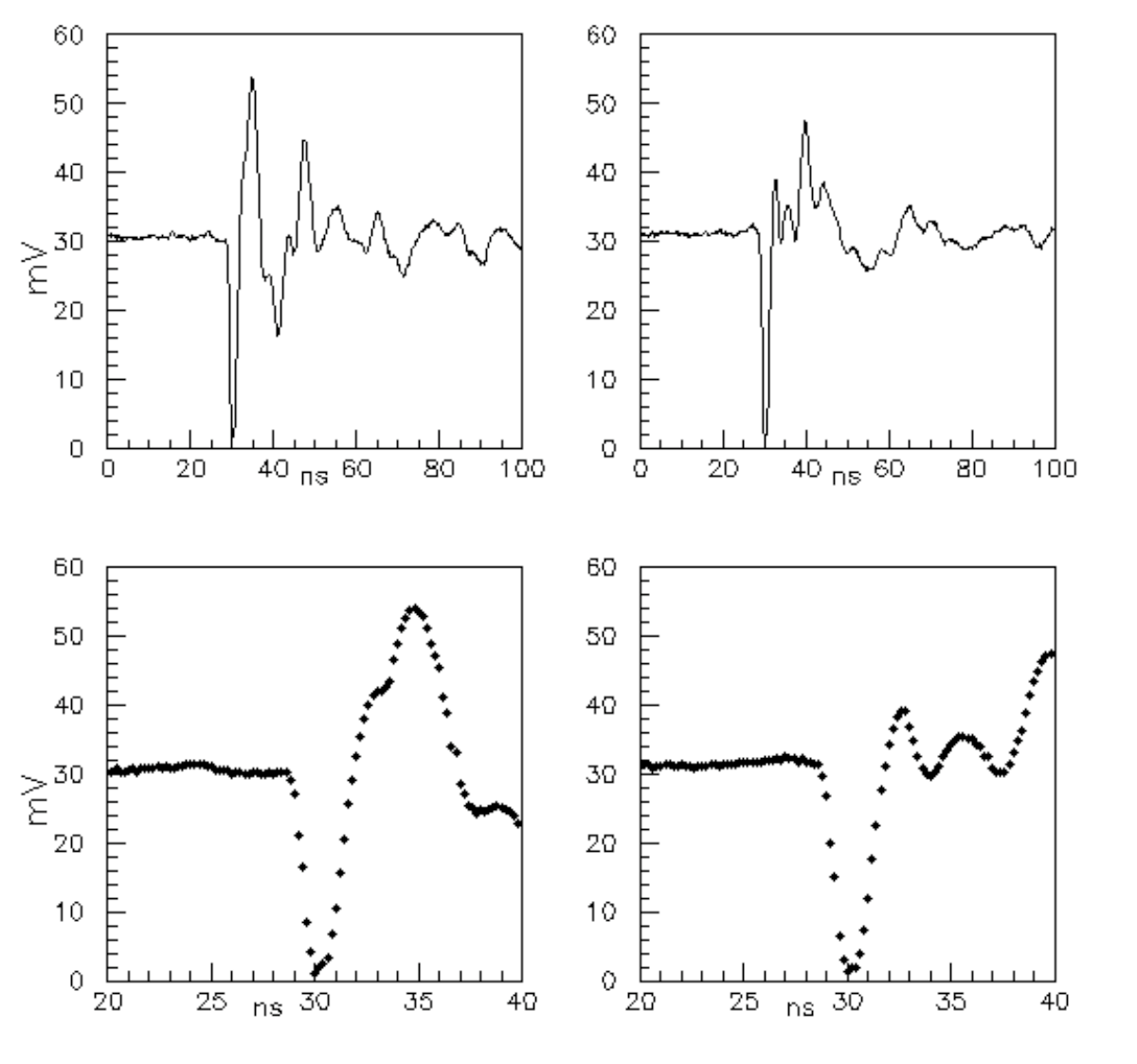}
\caption{Typical A5 (left) and A6 (right) sensor responses. The lower plots show in detail the
granularity of our time measurements (200 psec./bin).}
\label{fig4}
\end{figure}
The pulse shape which, unlike the current intensity, depends on the inductance,
capacitance and resistance (L, C, R) of the detectors.

The sensor response $V(t)$ for a step excitation $V_0$ can be written as:
\be
V(t) =  V_o \cdot e^{-\frac{R}{2L}t}sin(\omega t).
\label{curvateorica}
\ee
The natural frequency of the detectors is $\approx$ 250 MHz.
The voltage difference between the bar ends for the maximum value
of the Coulomb field, obtained suitably modifying  eqn.~\ref{dd}   
for a  finite longitudinal extent of the charge distribution (in our case, the
electron beam is $\approx$ 3 m. long) is:
\be
V^t_{max}=\eta \frac{\lambda}{2\pi \epsilon_o}ln\left(\frac{y+14.5~cm}{y}\right),
\label{volt}
\ee
where $\lambda$ is the charge per unit length of the incoming beam   
and $\eta$ is the sensor  calibration constant.
In the electric field calculations, the image charges
appearing on the flange as the beam exits the pipe have also been
included. However, as their effect decreases rapidly with the
distance from the flange, it is completely negligible in our
experiment (distance$\geqslant1\,$m).
The sensor calibration has been carried out using a known field
generated by a parallel plate capacitor.
We find experimentally $\eta=7.5\times 10^{-2}\pm\ 3\%$, however
due to various systematic effects we believe  our 
calibration to be good to $\approx$ 20\%, in absolute terms.

Assuming that the L.W. formula in eqn.\,\ref{volt} holds (which should  apply only if the uniform 
charge motion 
would last indefinitely and the charges generating the field would not be shielded by conductors)
we expect, in our typical beam operating conditions, pulse heights of the order of 10 mV  
out of our sensors. In the more realistic hypothesis that
the L.W. formula should be corrected to take into account the beam pipe shielding and the finite
lifetime for the charges uniform motion, as it is in our experiment, 
the expected amplitude, cfr.  fig.\ref{fig2}, would be 
of the order of few 
nanoVolt and hence unmeasurable.   

\begin{figure}[hbtp]
\begin{center}
\includegraphics[scale=0.45]{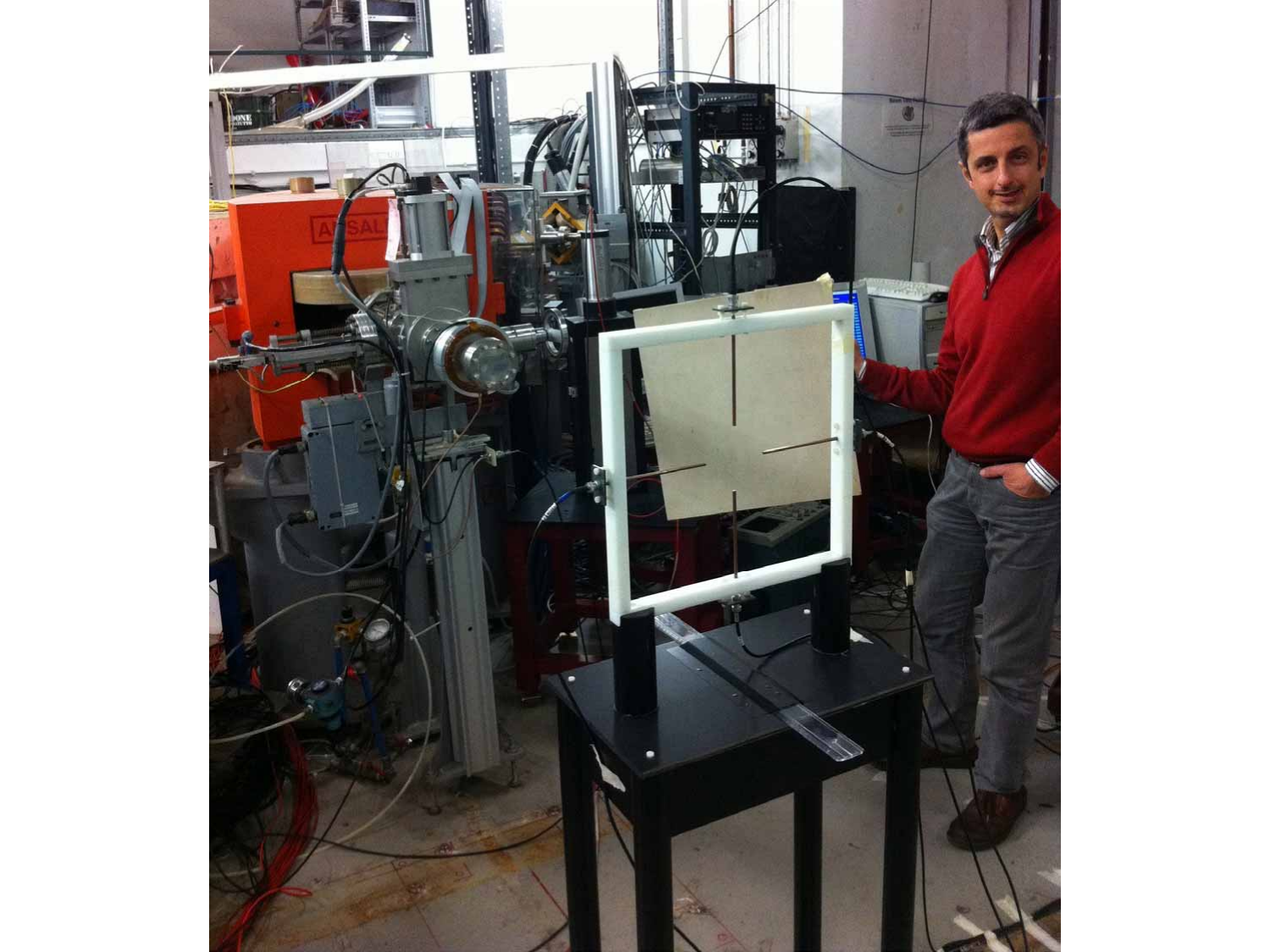}
\caption{
   A photograph of the beam pipe exit window and of the four reference
   sensors in the cross configuration.}
\label{fig5}
\end{center}
\end{figure}

We used  six sensors: four of them,  A1, A2, A3 and A4 in the following,
are located at a longitudinal distance of 92 cm from the beam exit
flange in a cross configuration, each
at  a transverse distance of 5 cm from the beam line
(cfr. fig.\,\ref{fig5}). The main purpose of 
these four sensors is  to provide reference  for the other two detectors 
A5 and A6 located through out the measurements  
at various longitudinal  and transverse coordinates along the beam trajectory.

\section{Measurements and data base }
Electron beams were delivered by BTF operators at a rate of few Hertz; data
were collected in different runs, identified by given longitudinal and
transverse position of the movable detectors (A5 and A6).

We collected a total of eighteen runs, spanning six transverse positions and 
three longitudinal positions of  A5 and A6 for a
total of about 15,000 triggers.
Through out the data taking, the references sensors (A1,A2,A3,A4) were 
left at the same location (92 cm. from the beam exit flange) 
in order to extract  a timing and amplitude
reference. As mentioned before, we collected data with the movable sensor at 
 172 cm, 329.5 cm and 552.5 cm  longitudinal distance  from the
beam exit flange. 
For each of the longitudinal positions we collected data on six transverse    
positions: 3, 5, 10, 20, 40 and 55 cm from the nominal 
beam line. 
For each run, A5 and A6 were positioned symmetrically with 
respect to the nominal beam line; spatial precision in the sensor positioning
was of the order of few mm in the longitudinal coordinate and 1\,mm in
the transverse one.

We define:
\be
{\bf Sn}=\frac{V_{max}\times 10^8}{N_{elec}},
\label{normamp}
\ee
where $V_{max}$ is the peak signal recorded by the SCA  and 
$N_{elec.}$ is the total number of electrons in the beam, as measured by the 
fast toroid. The factor $10^8$ in eqn.\ref{normamp} takes into account the typical
beam charge.  
\begin{figure}[hbtp]
\begin{center}
\vspace {-6cm}
\includegraphics[scale=0.6]{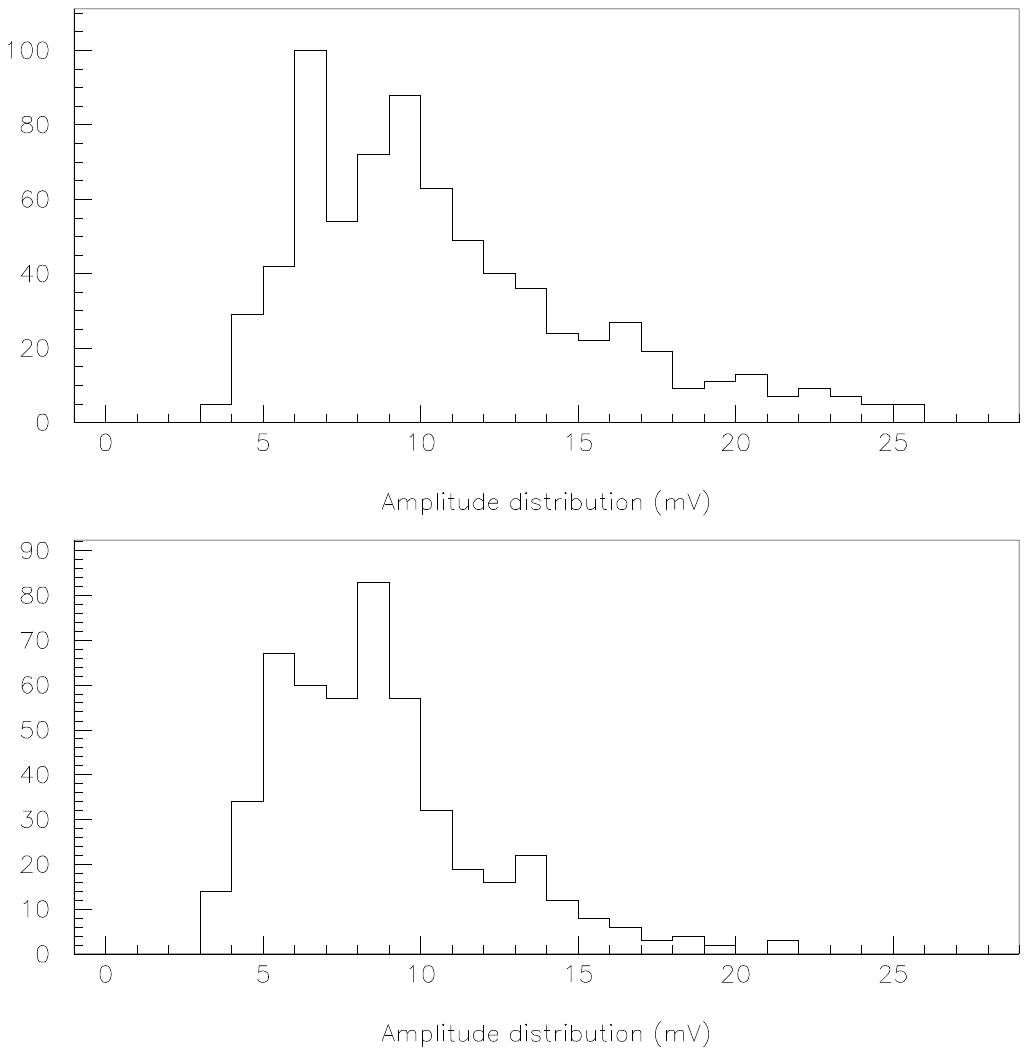}
\caption{Normalized amplitude {\bf Sn } 
(see text for details) for sensor A1 in
 two different runs: the above mentioned sensor was in the same operating 
conditions through out the different runs, taken minutes apart.}
\label{fig6}
\end{center}
\end{figure}
\vspace{6. cm}
As an example, the two plots in Fig. \ref{fig6} show the {\bf Sn} values for the reference
sensor A1, for two different runs. One issue common to all our measurements
stands out clearly: in the same experimental conditions (sensor position,
trigger timing, cable lengths, DAQ settings) the two distributions are  different.
 We attribute this difference to less than perfect
reliability in the beam delivery conditions (launch angles, total beam length,
charge distribution in the beam pulse length, stray magnetic fields, etc.),
over which we had little control.
 
Since our four reference sensors must provide normalization for the measurements
 taken by A5 and A6, our analysis proceeds as follows. We obtain, on a
run-by-run basis, the average amplitude of the four reference sensors, either
evaluating the medians, or fitting the distributions with Gaussian functions 
and taking the mean values.
 
Next we make the assumption that in any given series of runs under
study, different only for the position of the movable sensors A5 and A6, each
reference sensor, that is never moved, must always yield the same amplitude.
Since Fig. \ref{fig6} shows that this is not the case, we need to allow for some (uncontrolled)
 effect due to variation of beam parameters. We do so by enlarging
the errors on the reference amplitudes, originating from the Gaussian fits,
 by a rescaling factor. This factor is chosen by requesting
 that the reduced $\chi^2$ of the series be consistent with the hypothesis that all
amplitude measurements for any given sensor in the series have one common
value.
 
Once obtained the error rescaling factor, for a given sensor and run series,
we proceed to analyze the movable detectors by enlarging by the same factor
their own uncertainties. The amount of rescaling needed is of the order of 10;
the overall relative error on the run-averaged pulse height is typically $\approx$
10\%.

\subsection{Amplitude as a function of transverse distance}
As mentioned in the previous section, at each longitudinal position we collected data at six 
different transverse positions for  A5 and A6.
The requirements placed on data were: a lower cut on the beam charge 
($N_e > 0.5\times 10^8$) and upper cut on the baseline noise on the six detectors 
(noise $<$ 0.5 mV , where the typical r.m.s. noise was 0.15 mV).
We select in this way roughly 70\% of the recorded triggers.

Fig. \ref{fig7}, \ref{fig8} and \ref{fig9}  show normalized amplitudes 
{\bf Sn}  versus transverse 
distance obtained 
for the three different longitudinal positions. The 
displayed results are completely consistent with eqn.\ref{volt}, which gives
the voltage value expected in  case of a charge indefinitely moving with constant speed.
\begin{figure}[hbtp]
\begin{center}
\vspace{-2.0 cm}
\includegraphics[scale=0.5]{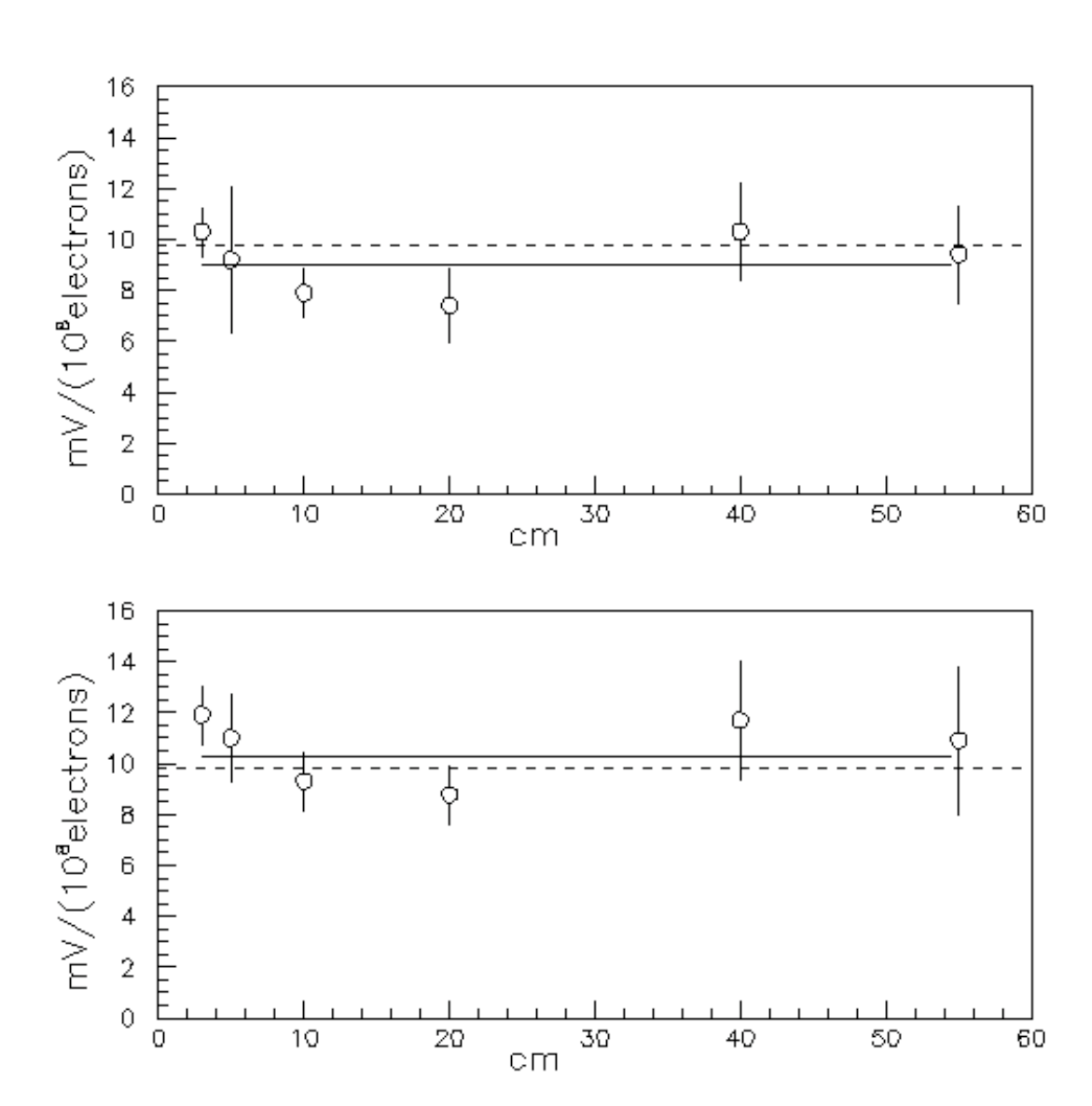}
\caption{ Comparison between measurements and predictions from eqn \ref{volt}:
normalized amplitudes {\bf Sn}  for sensors A1(upper)  and A3(lower) for
 $z_{A5,A6}$ =172.0 cm. The continuous line at 8.98$\pm$0.54 mV (upper) and 10.25$\pm$0.59 mV 
(lower) indicate the weighted average of our measurements.
The six measurements plotted refer to the transverse positions of sensor A5, A6. No dependence is 
expected (see text), as sensor A1, A3 were kept in the same operating conditions and at the
same locations. The dashed line indicates the nominal 
normalized value $V^t_{max}=9.78~mV$ of eqn.\ref{volt} for y=5 cm.
The agreement between measurements and prediction is remarkable  }
\label{fig7}
\end{center}
\end{figure}
\vspace{2.0 cm}
\begin{figure}[hbtp]
\begin{center}
\includegraphics[scale=0.5]{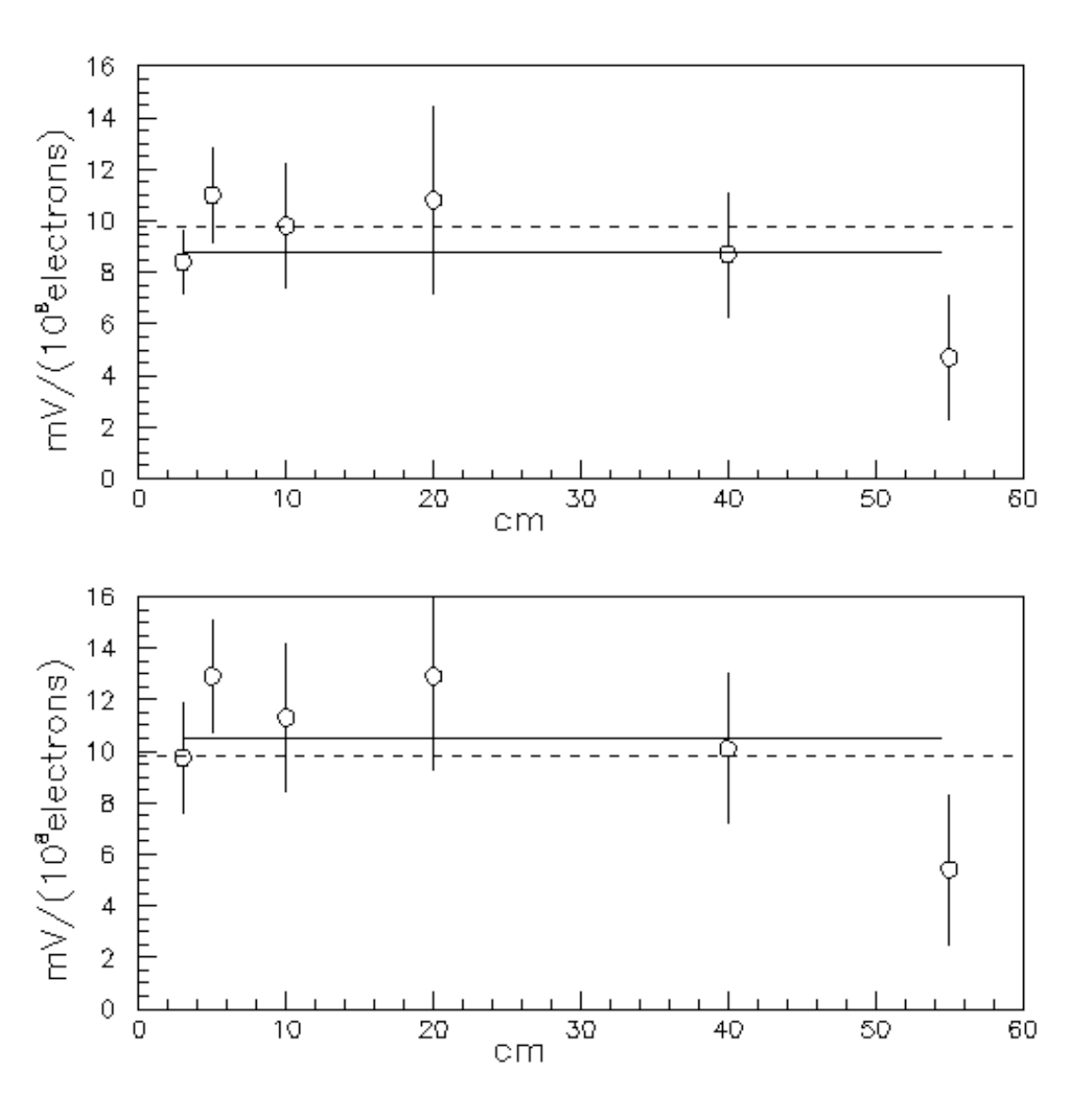}
\caption{The same plot as in  fig.\ref{fig7} at  $z_{A5,A6}$= 329.5 cm. The continuous 
lines at 8.80
$\pm$0.63 mV (upper) and 10.47$\pm$0.86 (lower) indicate the weighted average of our measurements.}
\label{fig8}
\end{center}
\end{figure}
\begin{figure}[hbtp]
\begin{center}
\includegraphics[scale=0.4]{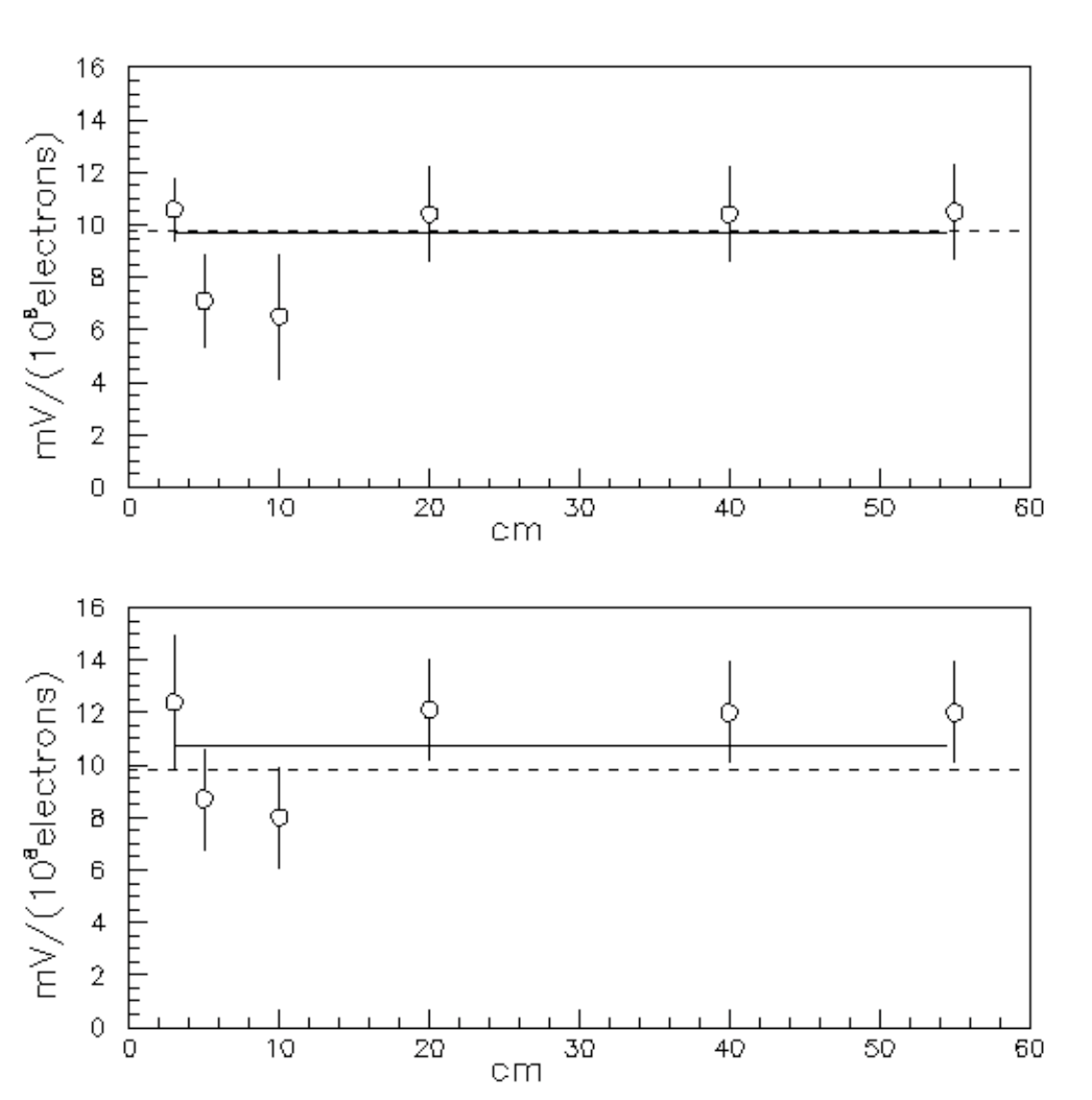}
\caption{The same plot as in fig.\ref{fig7}  at  $z_{A5,A6}$ = 552.5 cm. The continuous line  
at 9.67
$\pm$0.54 mV (upper) and 10.75$\pm$0.74 (lower) indicate the weighted average of our measurements.}
\label{fig9}
\end{center}
\end{figure}

The results shown in fig \ref{fig7}, \ref{fig8} and \ref{fig9} were obtained without
any normalization between measurements and L.W. theory. We stress again that the amplitude we measure
is many orders of magnitude higher than that one would expect from the {\it unshielded beam charge}.
Were we sensitive only to fields generated by the electron beam once they exited the beam pipe, 
our pulse height would have been, as mentioned in the previous paragraph, in the few nanoVolt 
range and then undetectable. 

In  figg. \ref{fig10},
 \ref{fig11} and \ref{fig12} we show the amplitude  ratios between sensors A5 and A1 (A6 and 
A3)  as a function of transverse distance from the beam line.
 Also in this case, data are completely consistent with the logarithmic behavior of  eqn \ref{volt}.
\begin{figure}[hbtp]
\begin{center}
\includegraphics[scale=0.5]{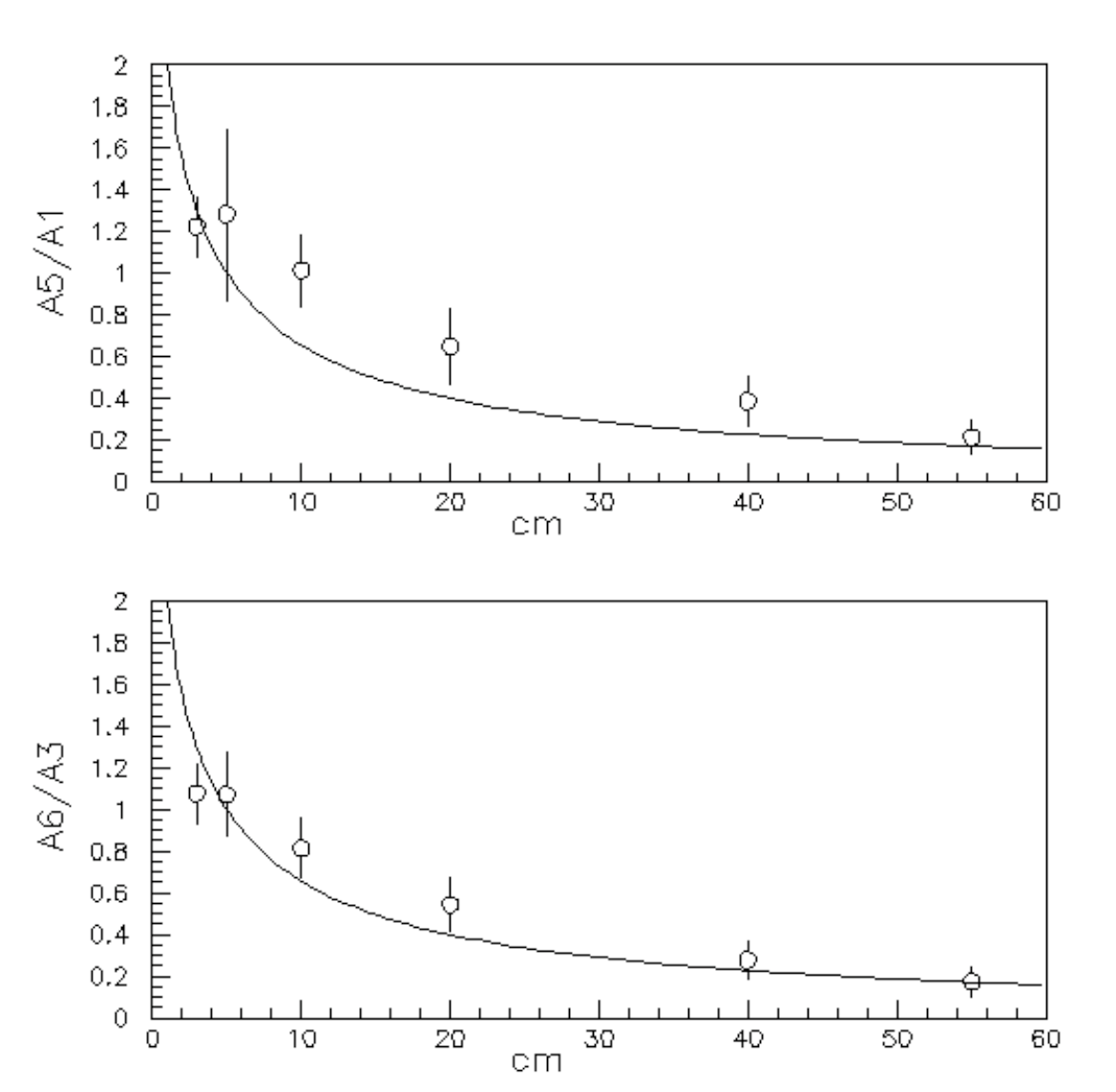}
\caption{Upper graph: the points  show the ratio   $V_{max}(A5) 
\over V_{max}(A1)$  at z$_{A5,A6}$=172.0 cm versus the transverse distance. 
Lower graph:  $V_{max}(A6) \over V_{max}(A3)$. The continuous lines represent
eqn. \ref{volt} for the depicted ratios. The two reduced $\chi^2$ are respectively
1.82 and 1.06. No fit has been performed on the data: the reduced $\chi^2$ has been
 evaluated from eqn \ref{volt} and the experimental data.}
\label{fig10}
\end{center}
\end{figure}
\begin{figure}[hbtp]
\begin{center}
\includegraphics[scale=0.5]{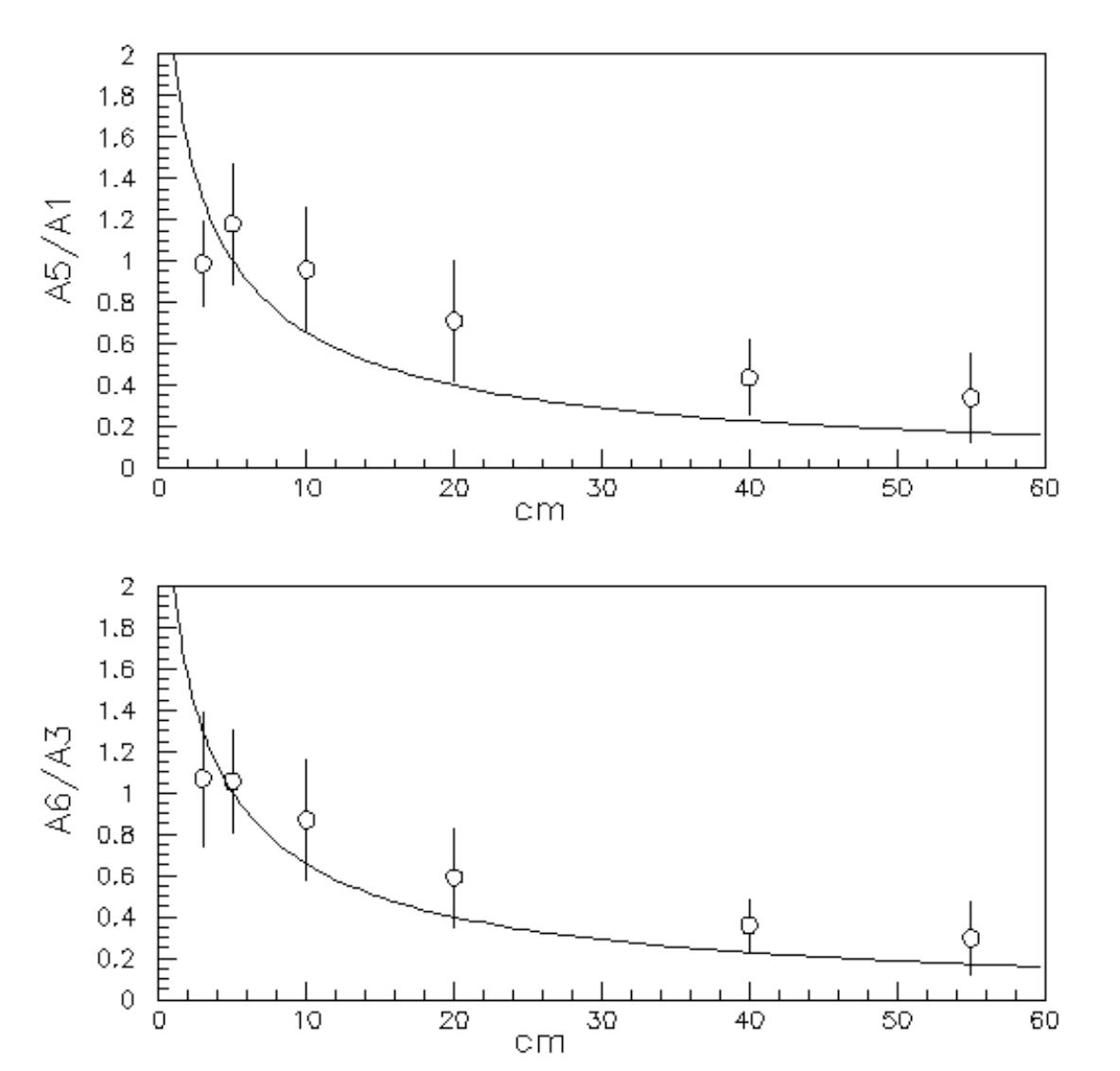}
\caption{The same plot as in fig.\ref{fig10} at  z$_{A5,A6}$=329.5 cm.
 The two reduced $\chi^2$ are respectively
1.36 and 0.66.}
\label{fig11}
\end{center}
\end{figure}
\begin{figure}[hbtp]
\begin{center}
\includegraphics[scale=0.5]{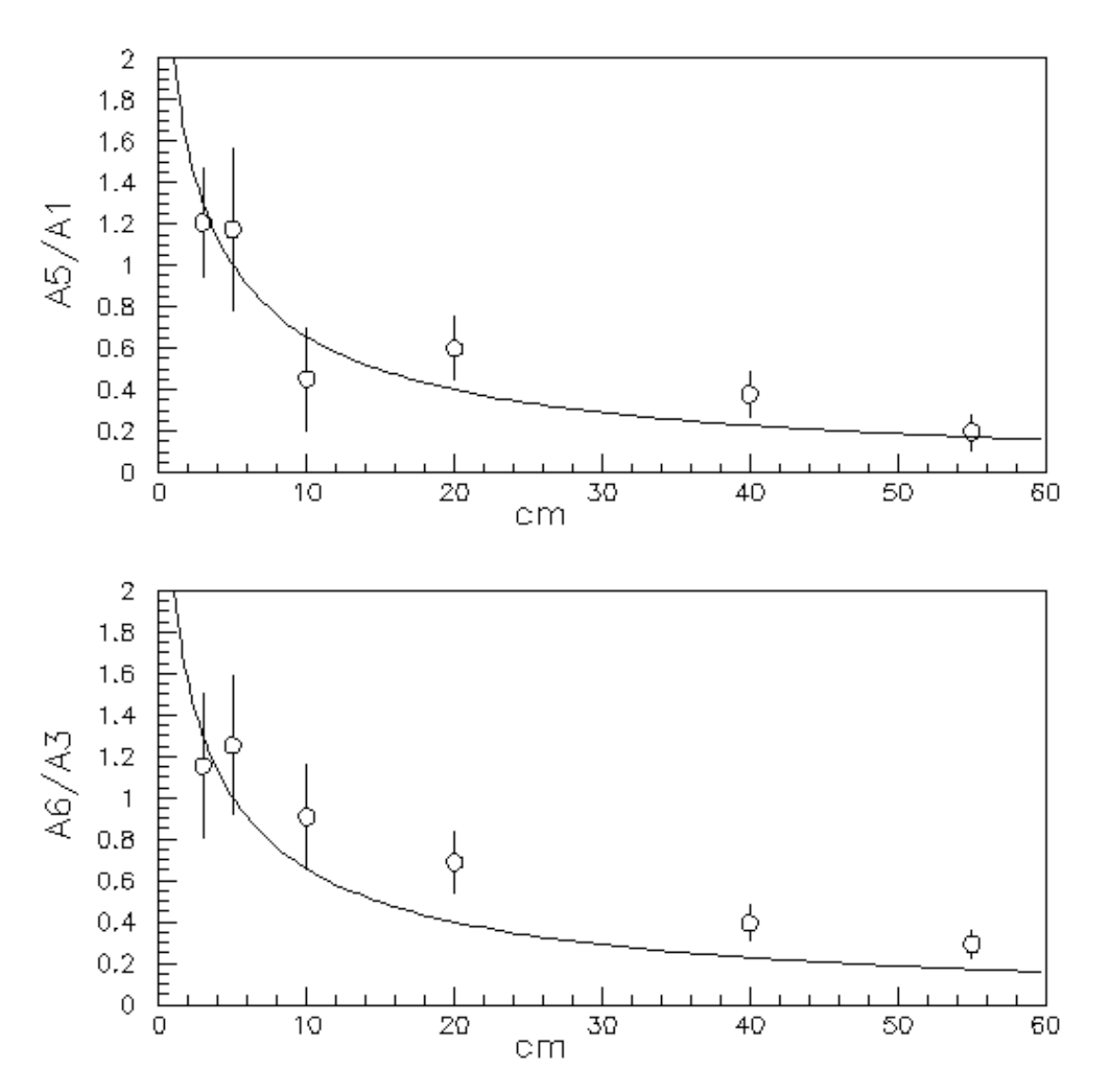}
\caption{The same plot as in fig.\ref{fig10}  at z$_{A5,A6}$=552.5 cm
 The two reduced $\chi^2$ are respectively
0.91 and 2.48.}
\label{fig12}
\end{center}
\end{figure}

\subsection{Timing measurements}
Our 200 psec/chn SCA  provides  timing for detector outputs, so that it is 
possible to detect both longitudinal  and transverse position-time correlations.
As a reminder we stress again that, in the hypothesis of stationary constant speed motion,
no time difference is expected as a function of  transverse distance, while different longitudinal 
positions should  exhibit delays consistent with particles traveling 
at $\gamma\approx$ 1000. 

Also in this case, we will have to rescale the errors yielded by standard procedures 
extracting central values from quasi Gaussian distributions; in this case we impose  
that, by symmetry, the time difference between A5 and A6 
be independent from the transverse distance between detector and nominal beam line. 
We then duplicate the 
procedure described  in the previous section requiring that time difference  between
A5 and A6 for each run  all come from a common value.
\begin{figure}[hbtp]
\begin{center}
\includegraphics[scale=0.6]{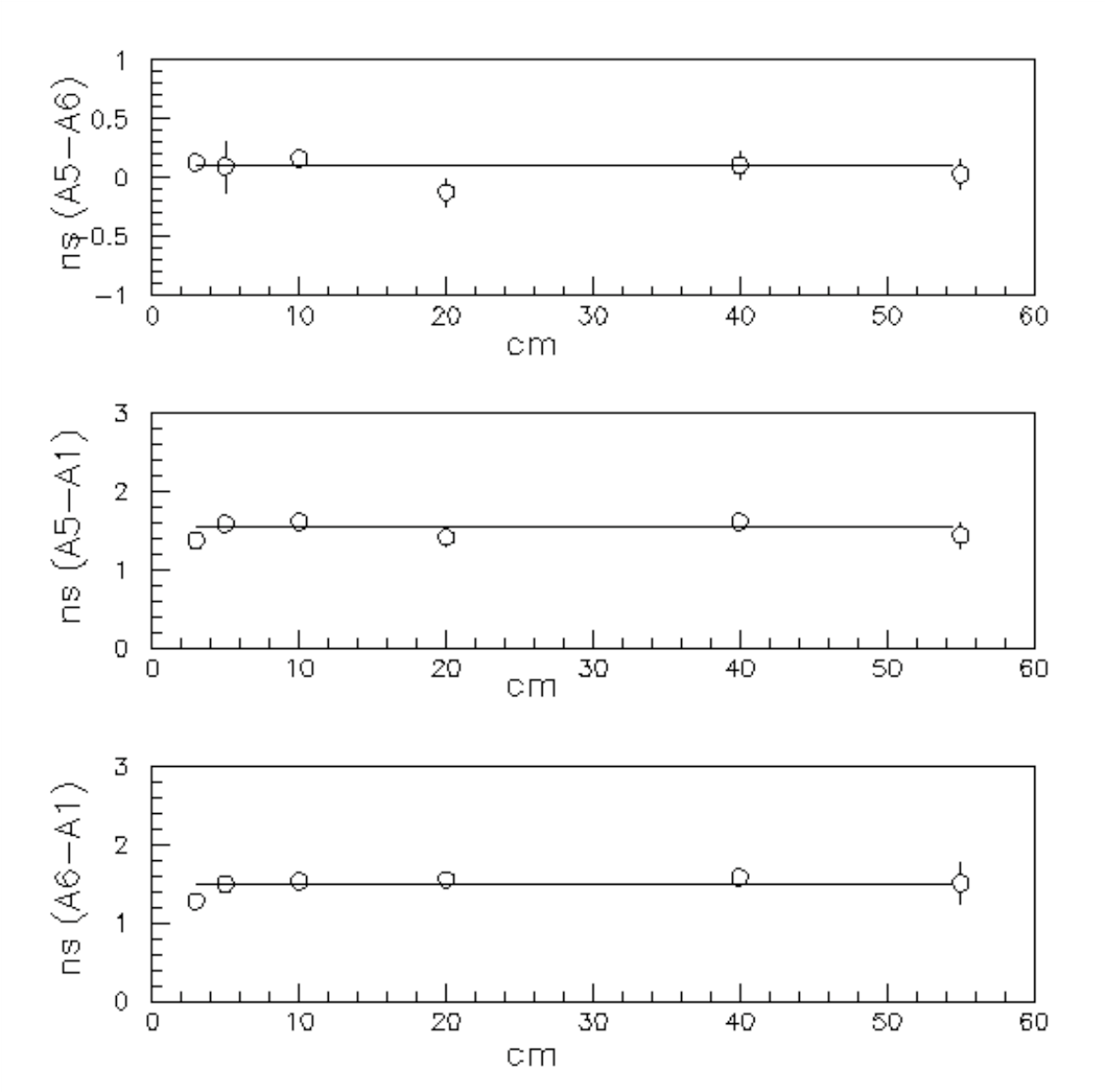}
\caption{Top graph: time difference between A5 and A6 versus the transverse distance. 
Errors have 
been rescaled according to the procedure 
described in the text.
Middle  and bottom plots: time difference between the movable sensors A5 and A6 (respectively) 
and one of the fixed sensors, A1. $z_{A5,A6}$ = 172 cm. The line at 1.549$\pm$0.036 nsec.(middle) and 
1.500$\pm$0.036 nsec (lower) indicate the weighted average of our measurements, once the 
reduced $\chi^2$ is rescaled according to the procedure described in the text
 (par. 4.2). The values obtained for the A5-A1, A6-A1 absolute delays  have 
to be corrected for the cables different lengths ($\approx$ 1 nsec.).}
\label{fig13}
\end{center}
\end{figure}

The upper graph of fig \ref{fig13} shows the time difference relative to 172 cm 
longitudinal
distance data. The amount of rescaling, in this case, is about a factor of 10 and
the overall resolution  on  time difference measurements is of the order of
50 psec.

The data show no time dependence of the sensor signal on transverse distance: the 
reduced $\chi^2$ for the hypothesis of a constant delay as  function of $y$ is always below 2 at each
longitudinal positions.
Furthermore, would one add a linear term  depending on transverse distance for  the sensor
time delay, the {\it inverse velocity} obtained would have a value smaller than 
$3\times 10^{-9}$ $sec\over m$ at 95\% confidence level.

We summarize the time distance correlations in Table \ref{time}, where the data 
obtained at the three different longitudinal positions are shown.

\begin{table}[hbtp]
\begin{center}
\caption{Timing measurements. The expected differences  are calculated for
500 MeV electrons. }
\vspace{3 mm}
\begin{tabular}{|c|ccc|}
\hline
longitudinal distances&expected&experimental&experimental\\
between two sensors [cm]&[ns]&a5 [ns]&a6 [ns]\\
\hline
(552.5-329.5) 223.0$\pm1.5$&$7.43\pm0.05$&$7.28\pm0.02$&$7.52\pm0.04$\\
(552.5-172.0) 380.5$\pm1.5$&$12.68\pm0.05$&$12.62\pm0.04$&$12.84\pm0.05$\\
(329.5-172.0) 157.5 $\pm15.$&$5.19\pm0.05$&$5.21\pm0.03$&$5.17\pm0.04$\\
\hline
\end{tabular}
\label{time}
\end{center}
\end{table}

\subsection{E.M. Backgrounds}
We preformed different tests in order to  ascertain that E.M. radiation coming from the interaction of 
the electron beam with its environment was not the original cause of our sensors' response.

With the beam steering system, we changed  the
launch angle in the experimental hall; varying  the current of the beam line magnet(s)
one can then predict the amplitude ratio of two detectors located right and left 
of the beam line, according to the calculated beam position at the sensors' longitudinal coordinates. 
Special runs were taken to this purpose and the results are completely
consistent with the expected horizontal beam displacement w.r.t. the nominal 
position.

Other E.M. phenomena are related to boundary crossings: as the beam travels between different media
({\it e.g.} the beam exit flange)
E.M. radiation can be generated. This, in turn, might mimic pulses we assume due to the interaction 
of the beam itself with our sensors.
The experimental situation can be schematized as a  Tamm \cite{tamm} problem:
 a beam of particles traveling  inside the 
vacuum pipe of the Linac, suddenly appears out of the end flange of the accelerator, moves with 
uniform velocity
through out the experimental hall ($\approx$ 7 m.) and disappears in the concrete wall of the hall.
A calculation of the expected effect, using the formulae reported in \cite{GarciaFernandez:2012yz} 
lent us confidence
that this background was not extremely relevant;  however,  in order to demonstrate that such 
a phenomenon does not contribute (or contributes very little) to our
 sensors' signal, we had a dedicated run during April 2014.

We collected data in two different modes:
\begin {enumerate}
\item {\it Calibration} runs in order to match the data collected during the 2012 campaign 
to the latest  (2014) runs.
\item  {\it Beam dump} runs  in which the electron beam was stopped in a 40 X$_0$ lead dump before 
reaching the vertical  detectors A5 and A6.
\end {enumerate}
The underlying idea was that data taken with the beam dump, would
yield the response of the A5, A6 
detectors in a no-beam situation thus allowing us to map the pulse height of our detectors when  
just backgrounds were present in the experimental hall. 
\begin{figure}[hbtp] 
\begin{center}
\includegraphics[scale=0.35]{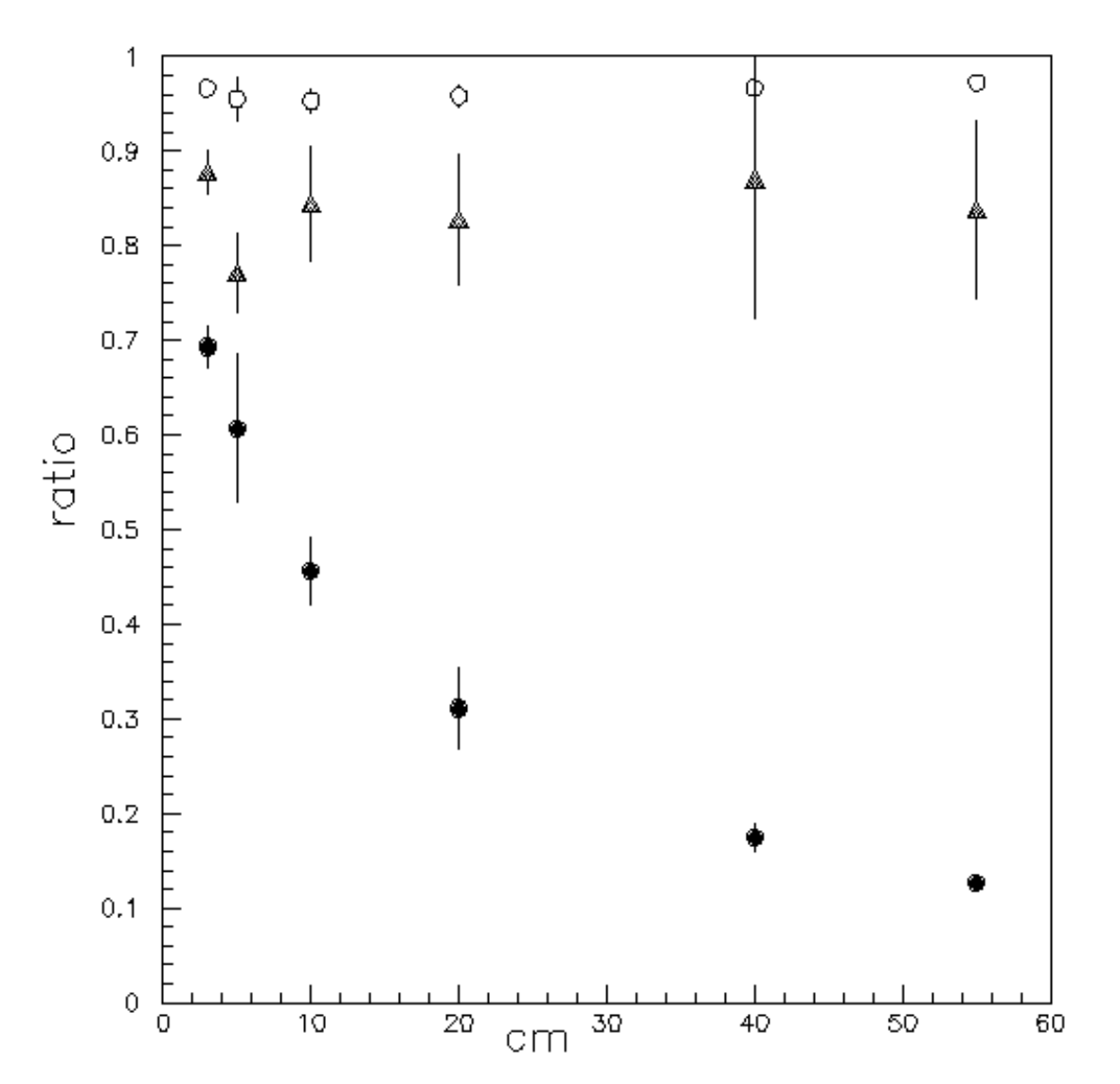}\\
\caption{Amplitude ratios vs transverse distance of sensors A5,A6  for 
{\it calibration runs}: open circles  A1 A3 ratio, triangles  A5 A6 ratio, 
 full dots   A5 A1 ratio. The plot is completely consistent with results
described in sec 4.1}
\label{fig14}
\end{center}
\end{figure}
From fig.\ref{fig14} one can infer that the main features of the previous (2012) measurements are 
retained 
in the (2014) latest run; small difference in the absolute values for the given ratios 
can be attributed to a less the perfect alignment of the two sets of detector on the beam line.
\begin{figure}[hbtp]
\begin{center}
\includegraphics[scale=0.35]{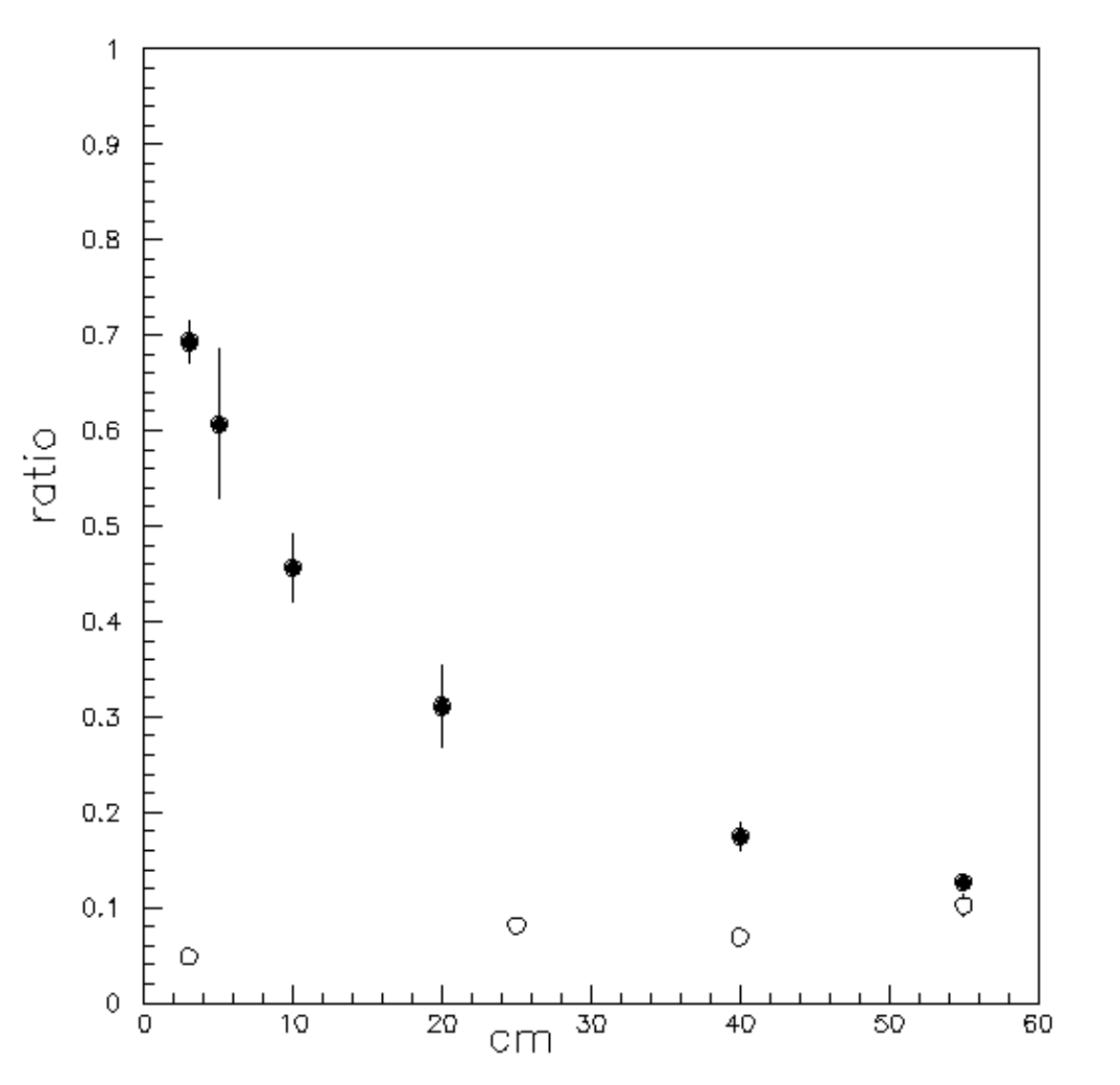}
\caption{Amplitude ratio for sensor A5 A1 vs (A5) transverse distance from the beam line: full dots
 calibration runs;  circles  runs taken with the beam dump inserted between A1...A4 and A5,
A6. Open circles data show unequivocally that any e.m. radiation originating 
either at the beam pipe boundaty or at the beam dump entrance does not score
on our detectors.} 
\label{fig15}
\end{center}
\end{figure}
Fig \ref{fig15} shows the comparison between data taken in the {\it calibration } mode and   
 the {\it beam dump} mode, that is when the 40 X$_0$ lead absorber is inserted between 
the A1...A4 and the A5,A6 sensors.
The vertical sensors  responses  are, with the beam dump in place,  
reduced by a factor $\approx$ 10, at 5 cm (transverse) distance, with practically no dependence on
(transverse) distance from the beam line. Such behavior lends itself to  the interpretation 
that the overall  amount of  E.M. background 
originating either at the transition flange or at the beam dump entrance is indeed  small 
w.r.t. the response obtained when beams unimpeded go  through the experimental hall.

  \section{Discussion}

With reference to table \ref{time}, we notice that the
longitudinal time differences are completely consistent with the
hypothesis of a beam traveling along the $z$ axis with a Lorentz factor $\gamma \approx 1000$.

Such an occurrence agrees
with the Li\'enard-Weichert model.  Retarded potentials, however,
predict that most of the {\it virtual photons}\,\cite{PanPhil} responsible for the
field detected at coordinates $z$ and $y$ be emitted several hundred meters before
the sensor positions and at different times according to the detectors transverse 
distances. 
Conversely, assuming that such virtual photons are
emitted in a physically meaningful region (between the
beam exit window and our detectors), the amplitude response of the sensors
should be several order of magnitude smaller than what is
being measured (cfr. Fig.\ref{fig2}).
Our result, obtained with a well definite set of boundary  conditions (longitudinal and
transverse distance between  beam line and sensors, details of the beam delivery to
the experimental hall etc.) matches precisely (within the experimental 
uncertainties) the expected value of the maximum field calculated 
according to  L.W. theory,  that is also the value calculated with Eq.\ref{e} when
 the beam is at the minimum distance from the sensor.

We again point out that the 
consistency of our measurement and eqn \ref{volt} has been obtained without any kind 
of normalization.

\section{Conclusions} 
The data we have discussed in the previous paragraphs led us to assume that 
the electric field of the electron beams acts on
our sensors only after the beam itself has exited the beam pipe and  that
Cerenkov and/or transition radiation effects are negligible. 

Our results agree with the prediction of L.W.formula, however if 
equation \ref{ee} is intended as if the fields were launched 
at an earlier time with respect to the sensors' response, 
such response would be orders of magnitudes smaller than the ones 
we measure.
The Feynman interpretation of the L.W. formula  for uniformly moving
charges does not show consistency with our experimental data.
Even if the steady state charge motion in our experiment lasted few tens of 
nanoseconds, our measurements indicate that everything behaves as if this
state lasted for much longer.

To summarize our finding in few words, one might  say that the data do
not agree with the most common interpretation 
of the Li\'enard-Weichert potential,
while seem to support the idea of a Coulomb field carried {\it
rigidly} by the electron beam.
    
\section{Acknowledgments}
We gratefully acknowledge Angelo Loinger for stimulating discussions, 
Ugo Amaldi for his interest in this research line and his advice,
Antonio Degasperis for cross-checking several calculations and Francesco Ronga
pointing out and discussing  Cerenkov and transition radiation effects.
We are indebted with Carlo Rovelli for his criticism and valuable suggestions.
We thank Giorgio Salvini for his interest in this research.

We  stress as well  the relevant contribution of our colleagues from the 
Frascati National Laboratory Accelerator Division and in particular
of Bruno Buonomo and Giovanni Mazzitelli.
Technical support of Giuseppe Mazzenga and Giuseppe Pileggi was also very valuable 
in the preparation and running of the experiment.
A special acknowledgment is in order for Paolo Valente, who in different stages of the
experiment, has provided, with his advice and ingenuity, solutions for various 
experimental challenges we faced.



\begin{thebibliography}{99}
\bibitem{eddi}A.Eddington \it Space, Time and Gravitation, \rm \\
Harper Torchbooks, pag 94 (1959)
\bibitem{laplace} Laplace, P., \it  Mechanique Celeste \rm, volumes published from 1799-1825, English
translation reprinted by Chelsea Publ., New York (1966) \\.
\bibitem{feyn}R. Feynman, R.B. Leighton, M.L. Sands, \it The Feynman Lectures on Physics, 
\rm  Addison-Wesley, Redwood City, vol. II, Chapters 21 and 26.2 (1989) \\
\bibitem{landau}L.D.Landau and E.M.Lifshitz \it The classical theory of fields, \rm pag.162, \\
Pergamon Press, Oxford (1971) \\
\bibitem{becker}R.Becker \it Teoria della elettricit\`{a}, \rm pagg. 73-77\\
Sansoni Ed. Scientifiche (1950) \\
\bibitem{jackson}J.D.Jackson \it Classical Electrodynamics, \rm pagg. 654-658\\
John Wiley \& Sons Inc. (1962,1975) \\
\bibitem{BTF}A. Ghigo, G. Mazzitelli, F. Sannibale, P.Valente, G. Vignola
{\it N.I.M.} A{\bf 515} 524-542  (2003) \\
\bibitem{tamm}  I.E. Tamm Jour. Phys. USSR {\bf 1} 439 1939 \\
\bibitem{GarciaFernandez:2012yz} 
  D.~Garcia-Fernandez, J.~Alvarez-Muniz, W.~R.~Carvalho, A.~Romero-Wolf and E.~Zas,
  ``Calculations of electric fields for radio detection of Ultra-High Energy particles,''
  Phys.\ Rev.\ D {\bf 87}, 023003 (2013)
  [arXiv:1210.1052 [astro-ph.HE]]\\  
\bibitem{PanPhil}  W. K. H.. Panofsky  and M. Philips,, \textit{Classical
    Electricity and Magnetism} \rm pagg. 350-351
Addison Wesley Publishing Co.(1955, 1962)\\
\end{thebibliography}
\end{document}